\documentclass[english,aps,preprint]{revtex4}
\usepackage[T1]{fontenc}
\usepackage[latin9]{inputenc}
\usepackage{array}
\usepackage{float}
\usepackage{multirow}
\usepackage{amsmath}
\usepackage{amssymb}
\usepackage{graphicx}

\makeatletter

\providecommand{\tabularnewline}{\\}

\@ifundefined{textcolor}{}
{%
 \definecolor{BLACK}{gray}{0}
 \definecolor{WHITE}{gray}{1}
 \definecolor{RED}{rgb}{1,0,0}
 \definecolor{GREEN}{rgb}{0,1,0}
 \definecolor{BLUE}{rgb}{0,0,1}
 \definecolor{CYAN}{cmyk}{1,0,0,0}
 \definecolor{MAGENTA}{cmyk}{0,1,0,0}
 \definecolor{YELLOW}{cmyk}{0,0,1,0}
}

\makeatother

\usepackage{babel}
\begin{document}

\title{Excited muon searches at the FCC based muon-hadron colliders}

\author{A. Caliskan}
\email{acaliskan@gumushane.edu.tr}

\affiliation{Gümü\c{s}hane University, Faculty of Engineering and Natural Sciences,
Department of Physics Engineering, 29100, Gümü\c{s}hane, Turkey}

\author{S. O. Kara}
\email{sokara@ohu.edu.tr}

\affiliation{Omer Halisdemir University, Bor Vocational School, 51240, Nigde,
Turkey}

\author{A. Ozansoy}
\email{aozansoy@science.ankara.edu.tr}

\affiliation{Ankara University, Faculty of Sciences, Department of Physics, 06100,
Tandogan, Ankara, Turkey}
\begin{abstract}
We study the excited muon production at the FCC-based muon-hadron
colliders. We give the excited muon decay widths and production cross-sections.
We deal with the $\mu p\rightarrow\mu^{\star}q\rightarrow\mu\gamma q$
process and we plot the transverse momentum and normalized pseudorapidity
distributions of final state particles to define the kinematical cuts
best suited for discovery. By using these cuts, we get the mass limits
for excited muons. It is shown that the discovery limits obtained
on the mass of $\mu^{\star}$ are 2.2, 5.8, and 7.5 TeV for muon energies
of $63$, $750$, and $1500$ GeV, respectively.
\end{abstract}
\maketitle

\section{\i ntroduction}

Discovery of the Higgs boson by ATLAS and CMS collaborations in 2012
\cite{key-1,key-2} has proved the accuracy and reliability of the
Standard Model (SM) of the particle physics. But, many questions about
dark matter, supersymmetric particles, extra dimensions, neutrino
masses, asymmetry between matter and anti-matter, existence of new
fundamental interactions, and fermion substucture are keeping their
mystery and waiting to be solved. Many theories beyond the SM (BSM)
have been proposed for these puzzling phenomena. Evidently, it is
necessary to perform the particle physics experiments in more powerful
colliders with higher energies and luminosities.

Compositeness is one of the BSM models that intend to solve the problem
of fermionic families replication, by introducing more fundamental
matter constituents called preons. Excited fermions are predicted
by preonic models and their existence would be a strong evidence for
fermion substructure \cite{key-3,key-4,key-5-1}. If known quarks
and leptons present composite structures, reasonable explanations
could be given for the still unanswered questions about the number
and replication of SM families and their mass hierarchy. The appearence
of excited states is an indisputable consequence of composite structure
of known fermions \cite{key-6,key-7-1,key-8-1,key-9-1}. In composite
models, SM fermions are considered as ground states of a rich and
heavier spectrum of excited states. Charged ($e^{\star},\mu^{\star},\tau^{\star}$)
and neutral ($\nu_{e}^{\star},\nu_{\mu}^{\star},\nu_{\tau}^{\star}$)
excited leptons come on the scene in the framework of composite models.
Excited leptons with spin-1/2 and weak-isospin-1/2 are considered
as the lowest radial and orbital excitations. Excited states with
higher spins also appear in composite models \cite{key-10-1,key-11,key-12,key-13,key-14-1}.

Considerable searches for the spin-1/2 charged and neutral excited
lepton signatures have been performed for the $e^{+}e^{-}$ and $ep$
colliders \cite{key-15-1,key-16-1,key-17-1,key-18-1}; $\gamma\gamma$
\cite{key-19,key-20,key-21,key-22-1} and $e\gamma$ \cite{key-14-1,key-23}
colliders; $pp$ \cite{key-24,key-25-1,key-26,key-27} and $p\overline{p}$
\cite{key-28,key-29,key-30} colliders. Production and decay properties
of spin-1/2 excited leptons in a left-right symmetric scenario are
studied in \cite{key-31}. Also, spin-3/2 excited leptons are studied
at various colliders in \cite{key-32,key-33,key-34,key-35,key-36,key-37,key-38}.

Excited electrons ($e^{\star}$) are extensively investigated in the
field of excited leptonic state studies. To perform a main comparison
it is necessary to study the other charged excited leptons ($\mu^{\star}$
and $\tau^{\star}$). In principle, $\mu^{\star}$ and $\tau^{\star}$
contributions would be differ from $e^{\star}$ contribution in the
mass and decay products of the SM leptons. 

The mass limit for excited spin-1/2 muons obtained from their pair
production ($e^{+}e^{-}\rightarrow\mu^{+\star}\mu^{-\star}$) by OPAL
collaboration at $\sqrt{s}=189-209$ GeV is $m_{\mu^{*}}>103.2$ GeV
\cite{key-39}. From single production ($pp\rightarrow\mu\mu^{\star}X$),
in events with three or more charged leptons at $\sqrt{s}=$8 TeV
including contact interactions in the $\mu^{\star}$ production and
decay mechanism, the ATLAS collaboration sets the mass limits as $m_{\mu^{*}}>3000$
GeV \cite{key-40}. Other studies on excited muon searches can be
found in \cite{key-41,key-42,key-43,key-44,key-45,key-46,key-47,key-48,key-49,key-50,key-51}.

Enormous efforts are being made for the research and development of
new particle colliders for the Large Hadron Collider (LHC) era and
post-LHC era. A staged approach will be taken into consideration for
the planning of these energy frontiers. The  first stage is low-energy
lepton colliders to make the precision measurements of the LHC discoveries.
These projects are the International Linear Collider (ILC) \cite{key-52}
with a center-of-mass energy of $\sqrt{s}=0.5$ TeV and low-energy
muon collider (a $\mu^{+}\mu^{-}$collider, shortly $\mu C$) \cite{key-53}.
Lepton-hadron collider projects would be considered as a second stage,
including an $ep$ collider under design, namely, Large Hadron Electron
Collider (LHeC) with $\sqrt{s}=1.3$ TeV (possibly upgraded to $\sqrt{s}=1.96$
TeV) \cite{key-54,key-55}, and a hypothetical $\mu p$ collider $\mu$-LHC
at this stage. The ILC with an increased center-of-mass energy ($\sqrt{s}=1$
TeV), the Compact Linear Collider (CLIC) \cite{key-56} with an optimal
center-of-mass energy of $3$ TeV, and the Plasma Wake-Field Accelarator-Linear
Collider project (PWFA-LC) \cite{key-57} are high-energy linear $e^{+}e^{-}$
colliders under consideration to be built after the LHC. On the side
of muon colliders, $\mu C$ with $\sqrt{s}$ up to 3 TeV is planned
as a high-energy muon collider \cite{key-53}.

The Future Circular Collider (FCC) \cite{key-58} project investigates
the various concepts of the circular colliders at CERN for the post-LHC
era. The FCC is proposed as the future $pp$ collider with $\sqrt{s}=100$
TeV and supported by European Union within the Horizon 2020 Framework
Programme for research and innovation. Besides the $pp$ option, it
is also being planned to include the $e^{+}e^{-}$ collider option
(TLEP or FCC-ee) \cite{key-59} and several $ep$ collider options
\cite{key-60,key-61}. 

Building a muon collider as dedicated $\mu$-ring tangential to the
FCC will give opportunity to handle multi-TeV scale $\mu p$ and $\mu A$
colliders \cite{key-62,key-63}. Assumed values for muon energy, center-of-mass
energy, and average instantaneous luminosity for different FCC-based
$\mu p$ collider options are given in Table I.

\begin{table}[H]
\caption{Main parameters of the FCC-based $\mu p$ collider}
\centering{}%
\begin{tabular}{|c|c|c|c|}
\hline 
Collider & $E_{\mu}$ (TeV) & $\sqrt{s}$ (TeV) & $L_{\mu p}(cm^{-2}s^{-1}$\tabularnewline
\hline 
\hline 
$\mu63$-FCC & 0.063 & 3.50 & $0.2\times10^{31}$\tabularnewline
\hline 
$\mu750$-FCC & 0.75 & 12.2 & $50\times10^{31}$\tabularnewline
\hline 
$\mu1500$-FCC & 1.5 & 17.3 & $50\times10^{31}$\tabularnewline
\hline 
\end{tabular}
\end{table}

Excited muon searches would provide complementary information for
the compositeness studies. This work is dedicated to search for excited
muons at future FCC-based muon-proton colliders. We introduce the
effective Lagrangian responsible for the gauge interactions of excited
muons and give their decay widths in Section II. Production cross-sections
and the analysis for the $\mu^{\star}\rightarrow\mu\gamma$ decay
mode are presented in Section III. We summarized our results in Section
IV. 

\section{effective lagrang\i an}

A spin-1/2 excited lepton is the lowest radial and orbital excitation
according to the classification by $SU(2)\times U(1)$ quantum numbers.
Interactions between excited spin-1/2 leptons and ordinary leptons
are of magnetic transition type \cite{key-15-1,key-16-1,key-64}.
The effective Lagrangian for the interaction between a spin-1/2 excited
lepton, a gauge boson ($V=\gamma,Z,W^{\pm}$), and the SM lepton is
given by

\begin{equation}
L=\frac{1}{2\Lambda}\overline{l_{R}^{*}}\sigma^{\mu\nu}\left[fg\frac{\vec{\tau}}{2}\centerdot\vec{W_{\mu\nu}}+f^{\prime}g^{\prime}\frac{Y}{2}B_{\mu\nu}\right]l_{L}+h.c.,
\end{equation}

where $\Lambda$ is the new physics scale, $W_{\mu\nu}$ and $B_{\mu\nu}$
are the field strength tensors, $\vec{\tau}$ denotes the Pauli matrices,
$Y$ is the hypercharge, $g$ and $g^{\prime}$ are the gauge couplings,
and $f$ and $f^{\prime}$ are the scaling factors for the gauge couplings
of $SU(2)$ and $U(1)$; $\sigma^{\mu\nu}=i(\gamma^{\mu}\gamma^{\nu}-\gamma^{\nu}\gamma^{\mu})/2$
with $\gamma^{\mu}$ being the Dirac matrices. An excited lepton has
three possible decay modes: radiative decay $l^{\star}\rightarrow l\gamma$,
neutral weak decay $l^{\star}\rightarrow lZ$ , and charged weak decay
$l^{\star}\rightarrow\nu W$. Neglecting the SM lepton mass, we find
the decay width of excited leptons as

\begin{equation}
\Gamma(l^{\star}\rightarrow lV)=\frac{\alpha m^{\star3}}{4\Lambda^{2}}f_{V}^{2}(1-\frac{m_{V}^{2}}{m^{\star2}})^{2}(1+\frac{m_{V}^{2}}{2m^{\star2}}),
\end{equation}

where $f_{V}$ is the new electroweak coupling parameter corresponding
to the gauge boson $V$, and $f_{\gamma}=-(f+f^{\prime})/2,$ $f_{Z}=(-fcot\theta_{W}+f^{\prime}tan\theta_{W})/2,$
and $f_{W}=f/\sqrt{2}sin\theta_{W}$; $\theta_{W}$ is the weak mixing
angle, and $m_{V}$ is the mass of the gauge boson, and $m^{*}$ is
the mass of the excited lepton. Total decay widths of excited leptons
for $\Lambda=m^{\star}$ and $\Lambda=100$ TeV are given in Figure
1. 

\begin{figure}[H]
\begin{centering}
\includegraphics{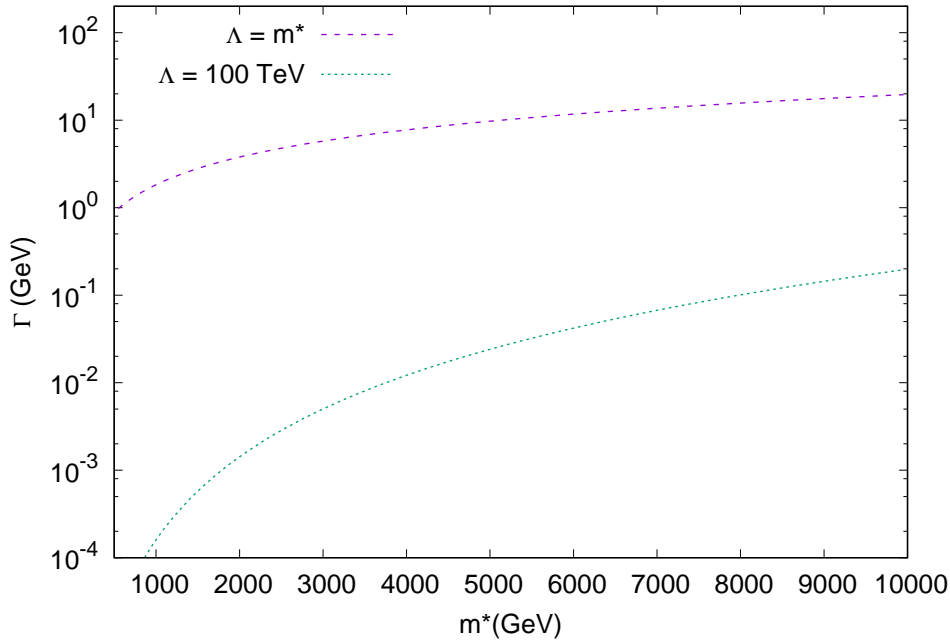}
\par\end{centering}
\caption{Decay width of excited leptons for $\varLambda=m^{*}$ and $\varLambda=100$
TeV.}
\end{figure}

\section{excited muon production at $\mu p$ colliders}

The FCC-based $\mu p$ colliders will provide the potential reach
for excited muon searches through the $\mu p\rightarrow\mu^{\star}X$
process. Feynman diagrams for the subprocesses $\mu q(\bar{q})\rightarrow\mu^{\star}q(\bar{q)}$
are shown in Figure 2. We implemented excited muon interaction vertices
in high-energy physics simulation programme CALCHEP \cite{key-65,key-66,key-67}
and used it in our calculations. 

\begin{figure}[H]
\begin{centering}
\includegraphics[scale=1.2]{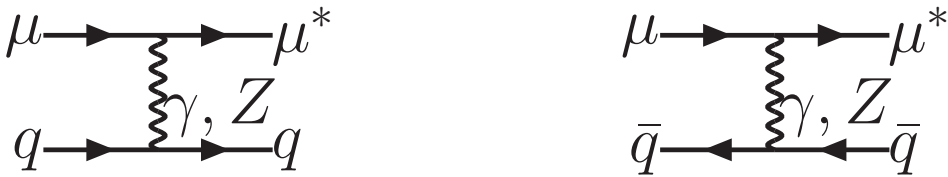}
\par\end{centering}
\centering{}\caption{Leading-order Feynman diagrams for the $\mu^{\star}$ production at
$\mu p$ collider.}
 
\end{figure}

Total cross-section for the process $\mu p\rightarrow\mu^{\star}X$
as a function of the excited muon mass is shown in Figure 3. We used
the CTEQ6L parton distribution function in our calculations.

\begin{figure}[H]
\begin{centering}
\includegraphics[scale=0.65]{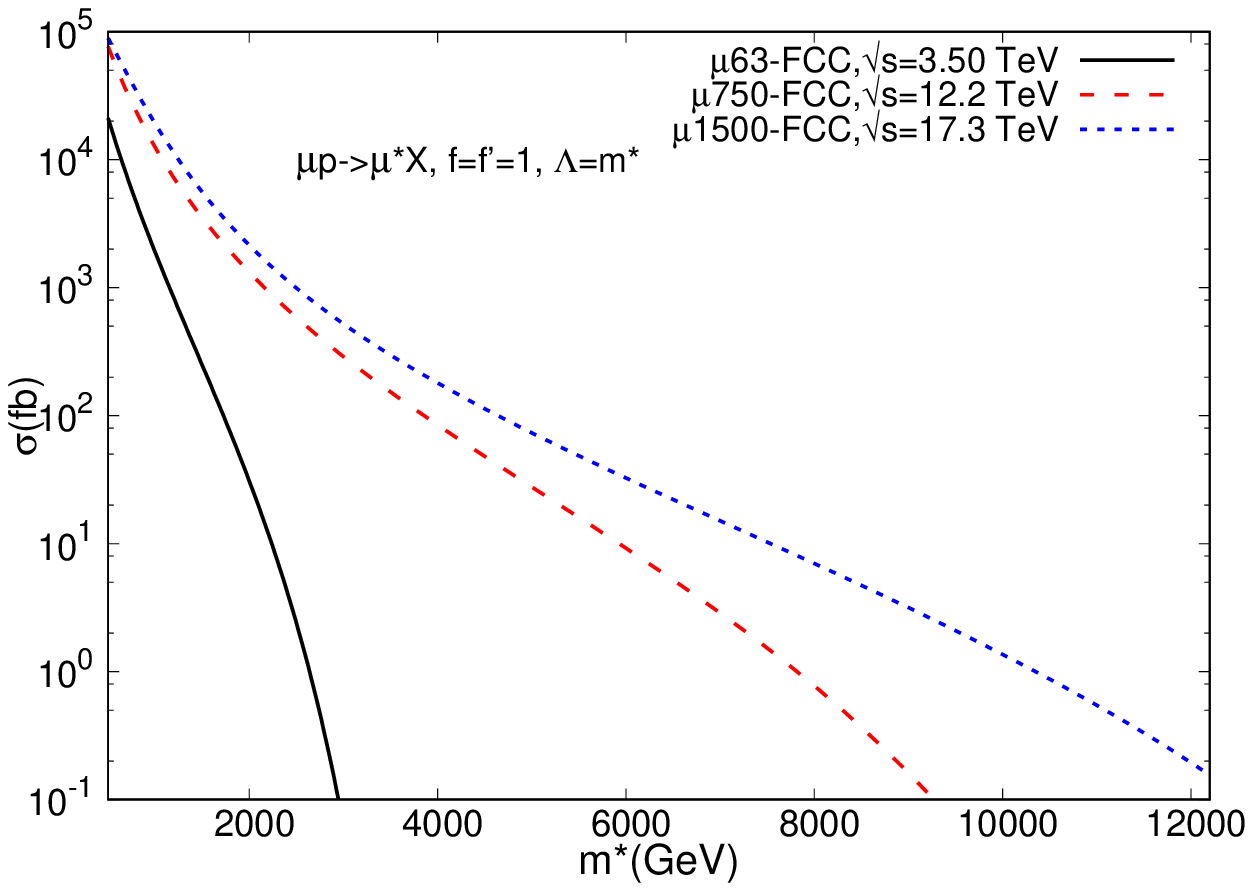}\includegraphics[scale=0.65]{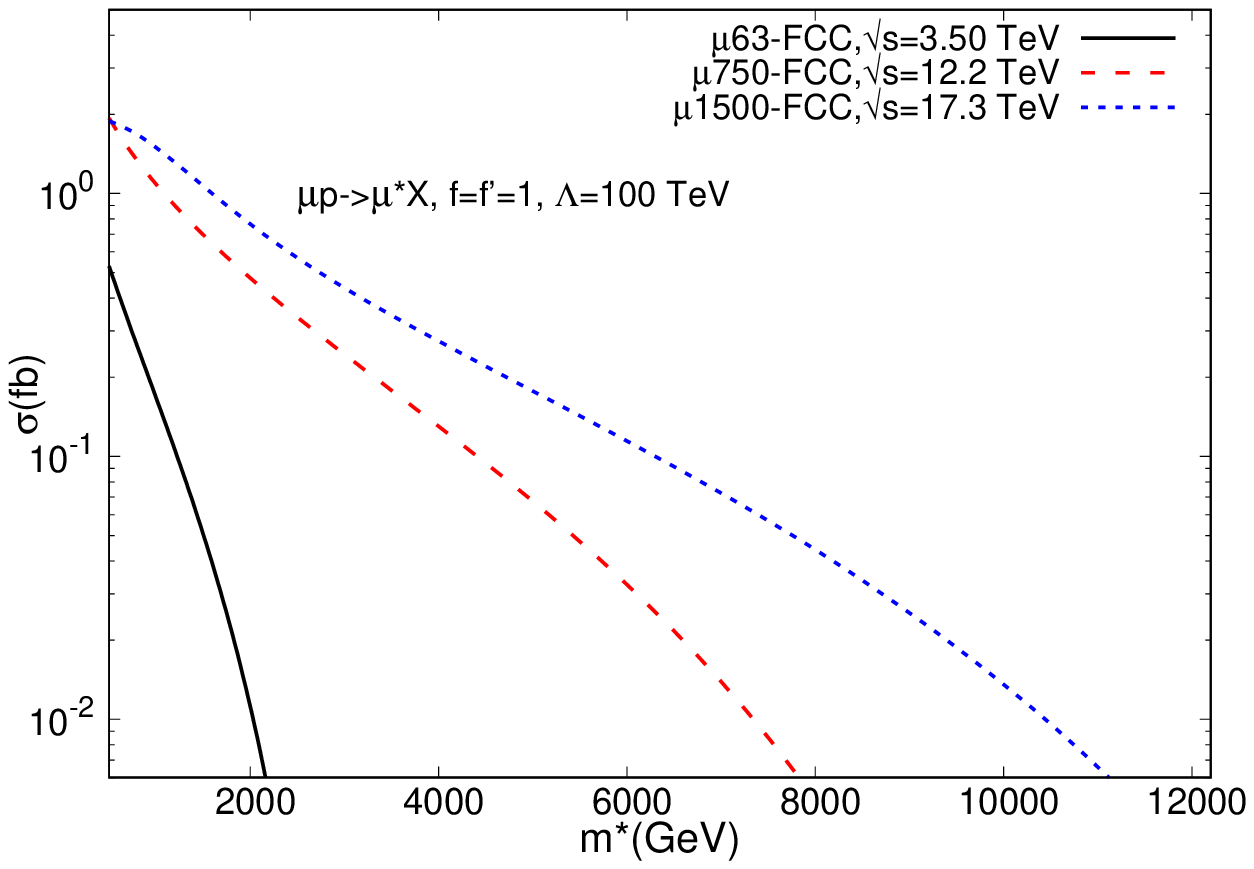}
\par\end{centering}
\caption{Total cross-section as a function of the excited muon mass for the
$\mu p$ colliders with various center-of-mass energies for $\varLambda$=$m^{*}$
(left) and $\varLambda$=$100$ (right) TeV, respectively.}
\end{figure}

For the analysis we take into account the $\mu\gamma$ decay mode
of the $\mu^{\star}$. We deal with the process $\mu p\rightarrow\mu^{\star}X\rightarrow\mu\gamma X$
(subprocess $\mu q(\bar{q})\rightarrow\mu\gamma q(\bar{q)}$) and
impose generic cuts, $p_{T}>20$ GeV, for the final state muon, photon
and jets.

Standard Model cross-sections after the application of the generic
cuts are $\sigma_{B}=24.51$ pb, $\sigma_{B}=89.69$ pb, and $\sigma_{B}=122.43$
pb for $\sqrt{s}=3.50,$ $12.2$ and $17.3$ TeV, respectively. We
show the transverse momentum distributions in Figure 4 (for $\mu63$-FCC),
in Figure 6 (for $\mu750$-FCC), and in Figure 8 (for $\mu1500$-FCC);
the normalized pseudorapidity distributions are in Figure 5 (for $\mu63$-FCC),
in Figure 7 (for $\mu750$-FCC), and in Figure 9 (for $\mu1500$-FCC).
We choose $f=f^{\prime}=1$ and $\Lambda=m_{\mu}^{\star}$ in our
calculations. As it is seen from Figures 4, 6 and 8 excited muons
carry high transverse momentum and these distributions show a peak
around $m_{\mu^{*}}/2$. Also, normalized pseudorapidity distributions
are so asymmetric. Since pseudorapidity is defined to be $\eta=-ln(tan(\theta/2))$,
where $\theta$ is the polar angle, it is concluded that excited muons
are produced mostly in the backward direction.

\begin{figure}[H]
\includegraphics[scale=0.65]{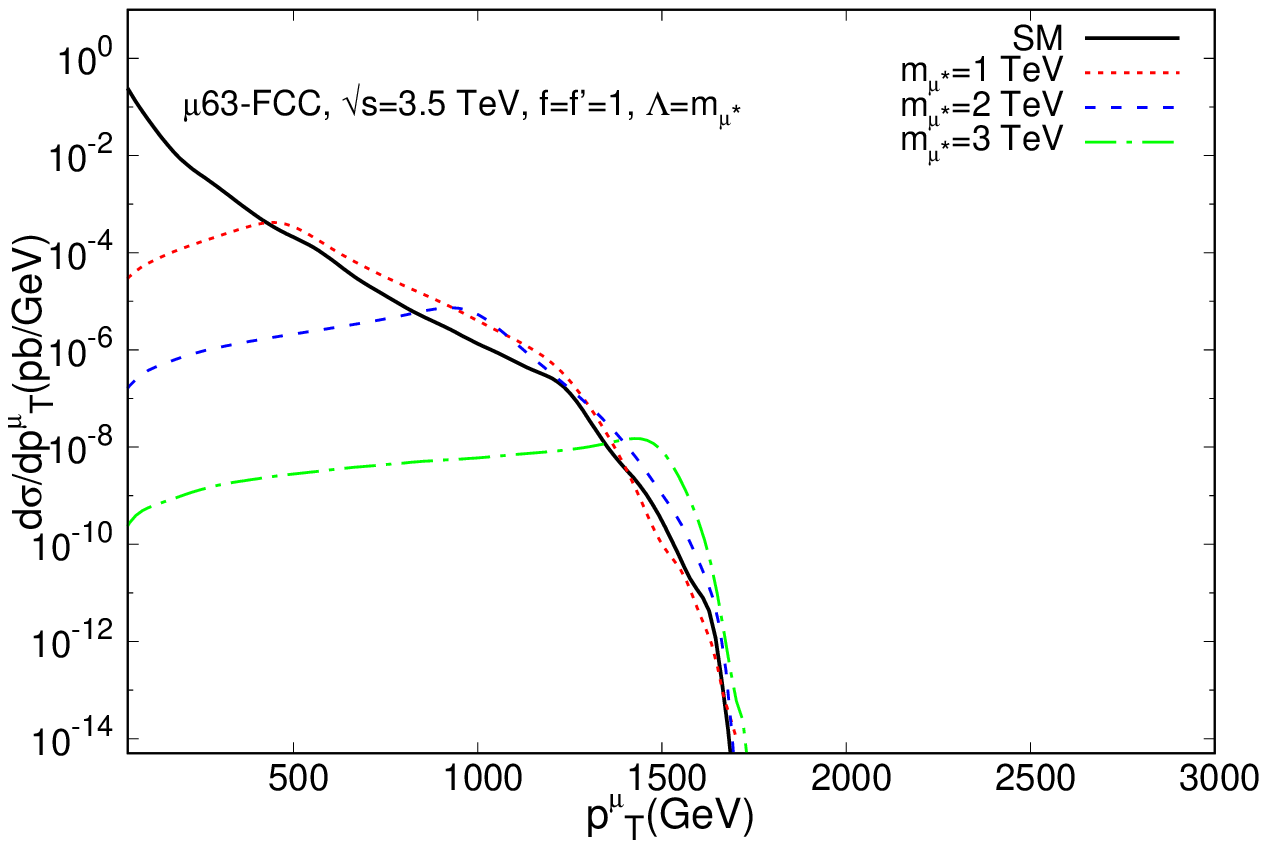}\includegraphics[scale=0.65]{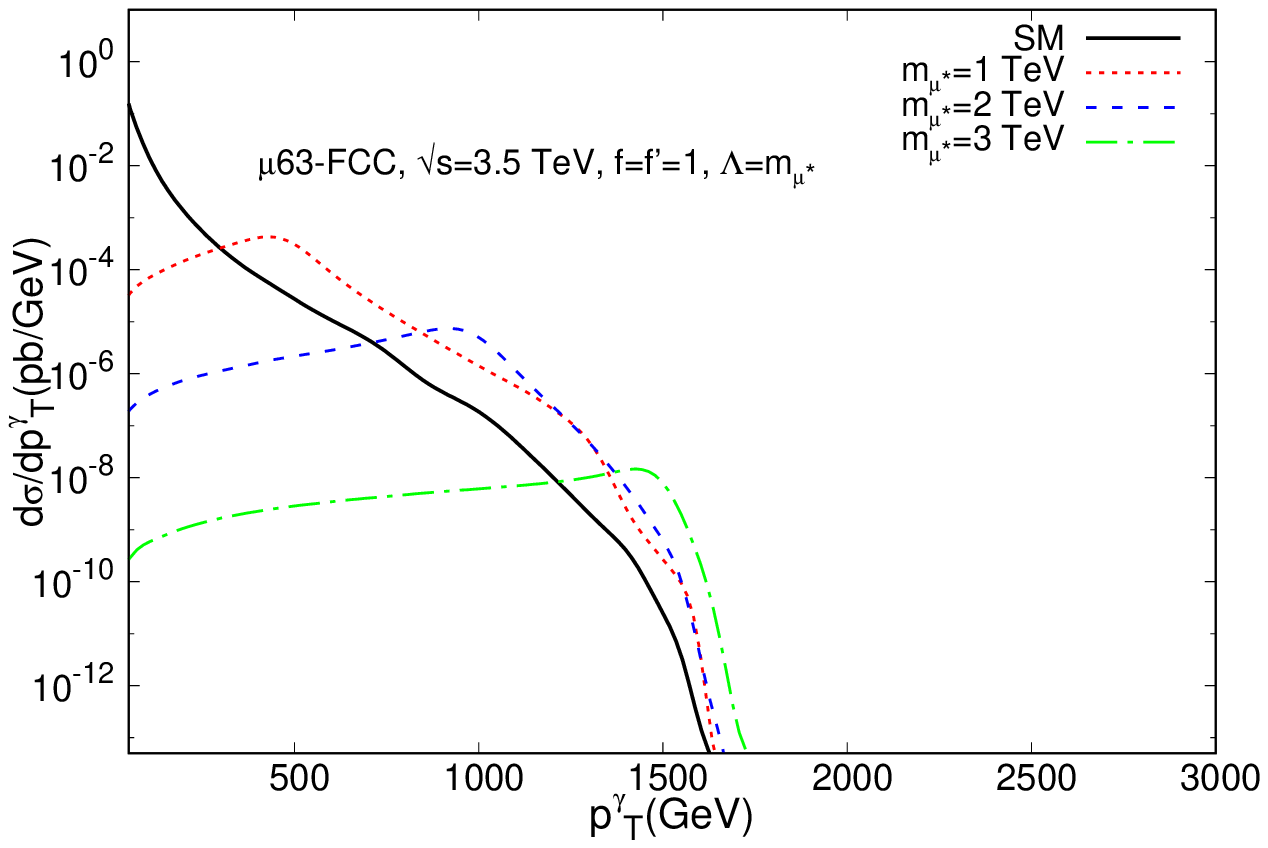}

\caption{Muon (left) and photon (right) $p_{T}$ distributions for the $\mu63$-FCC.}
\end{figure}

\begin{figure}[H]
\includegraphics[scale=0.65]{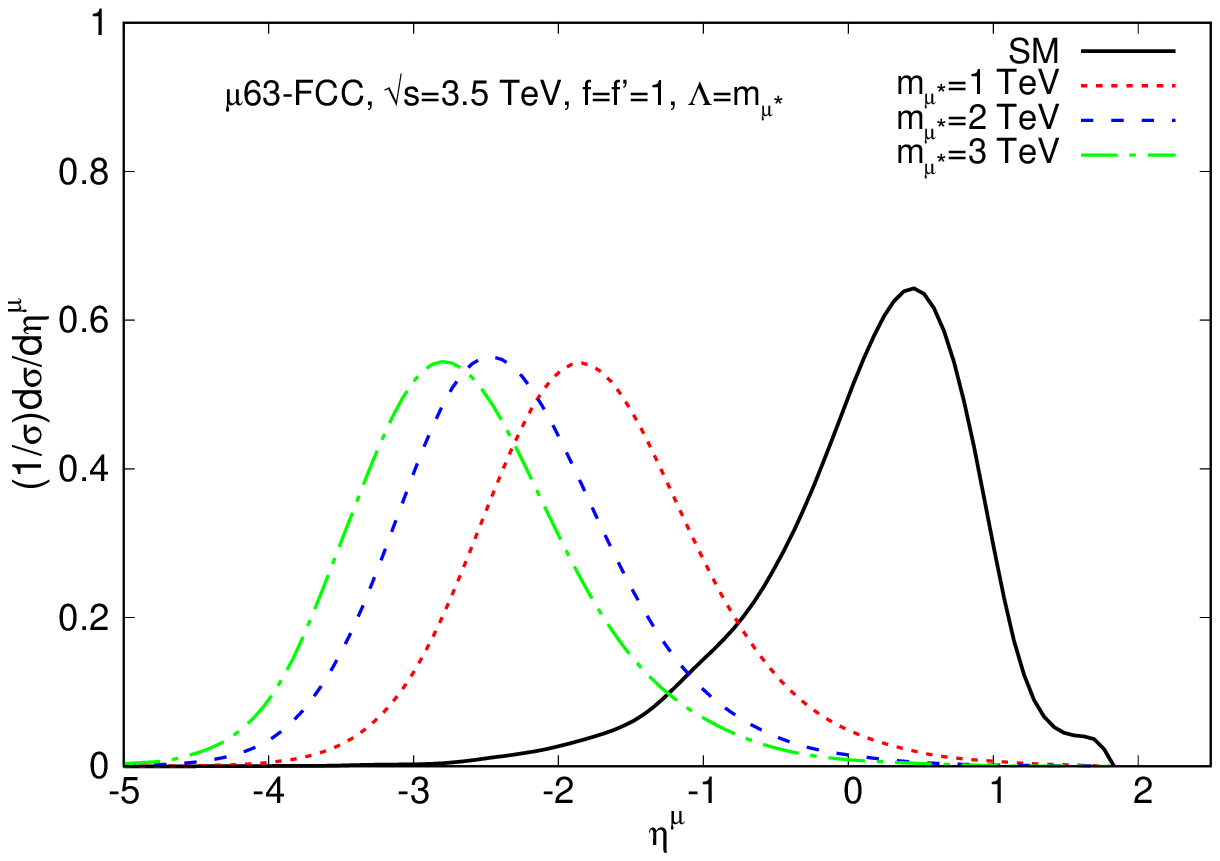}\includegraphics[scale=0.65]{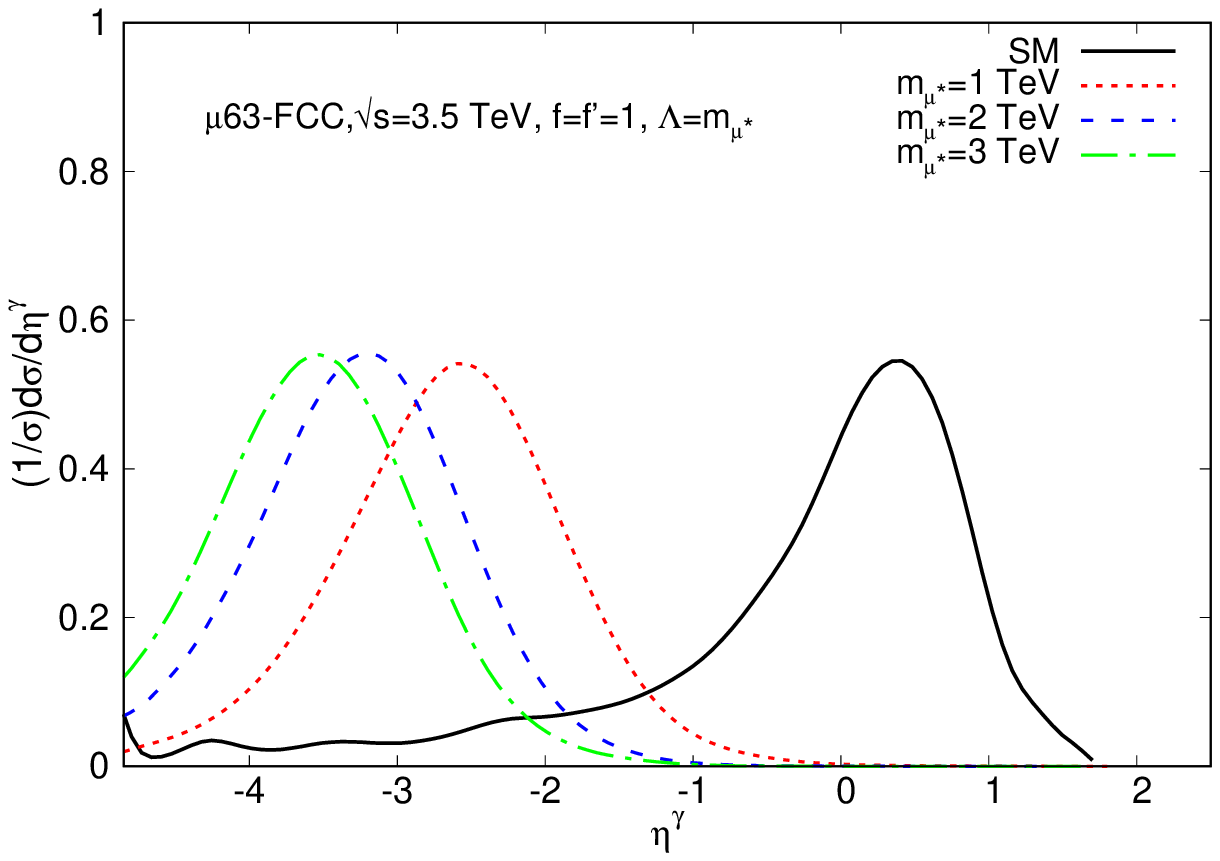}

\caption{Muon (left) and photon (right) normalized $\eta$ distributions for
the $\mu63$-FCC.}
\end{figure}

\begin{figure}[H]
\includegraphics[scale=0.65]{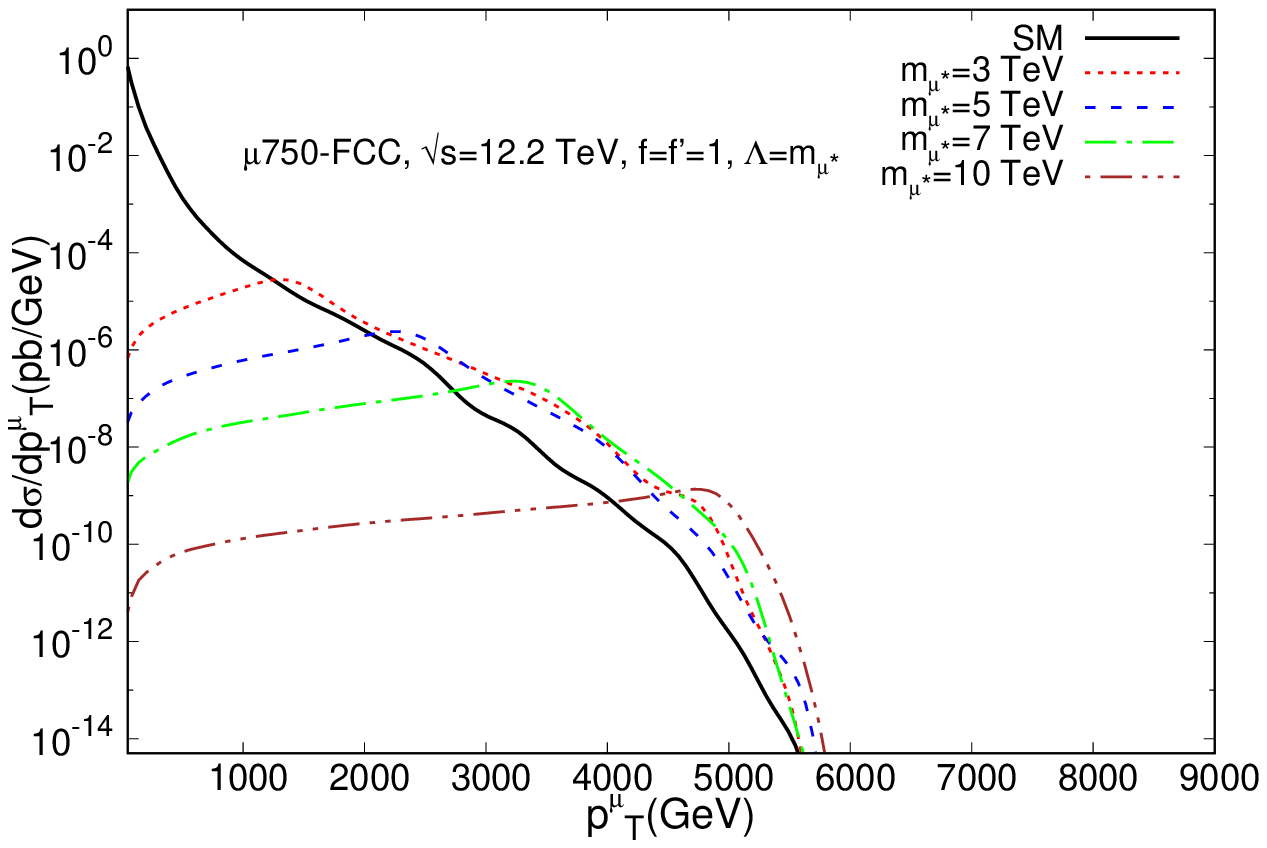}\includegraphics[scale=0.65]{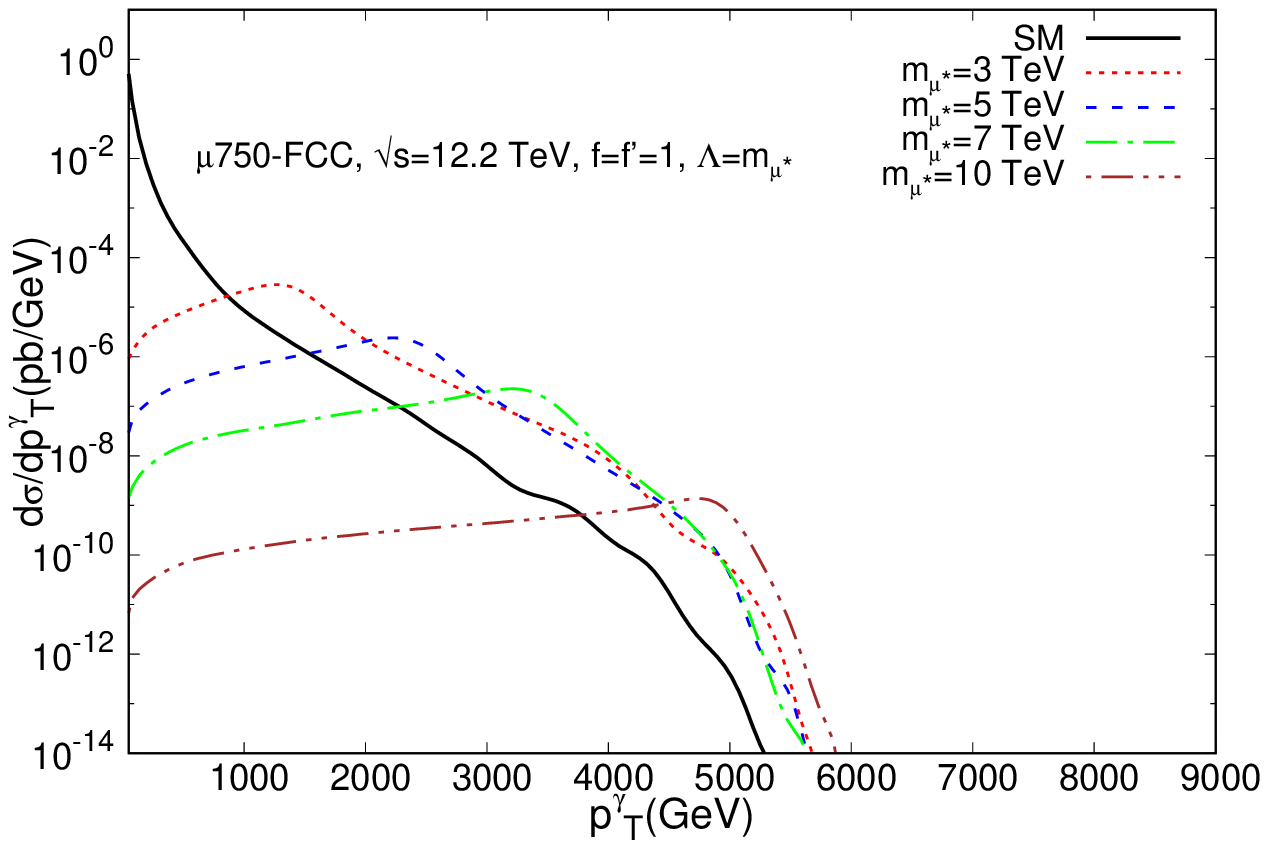}\caption{Muon (left) and photon (right) $p_{T}$ distributions for the $\mu750$-FCC.}
\end{figure}

\begin{figure}[H]
\includegraphics[scale=0.65]{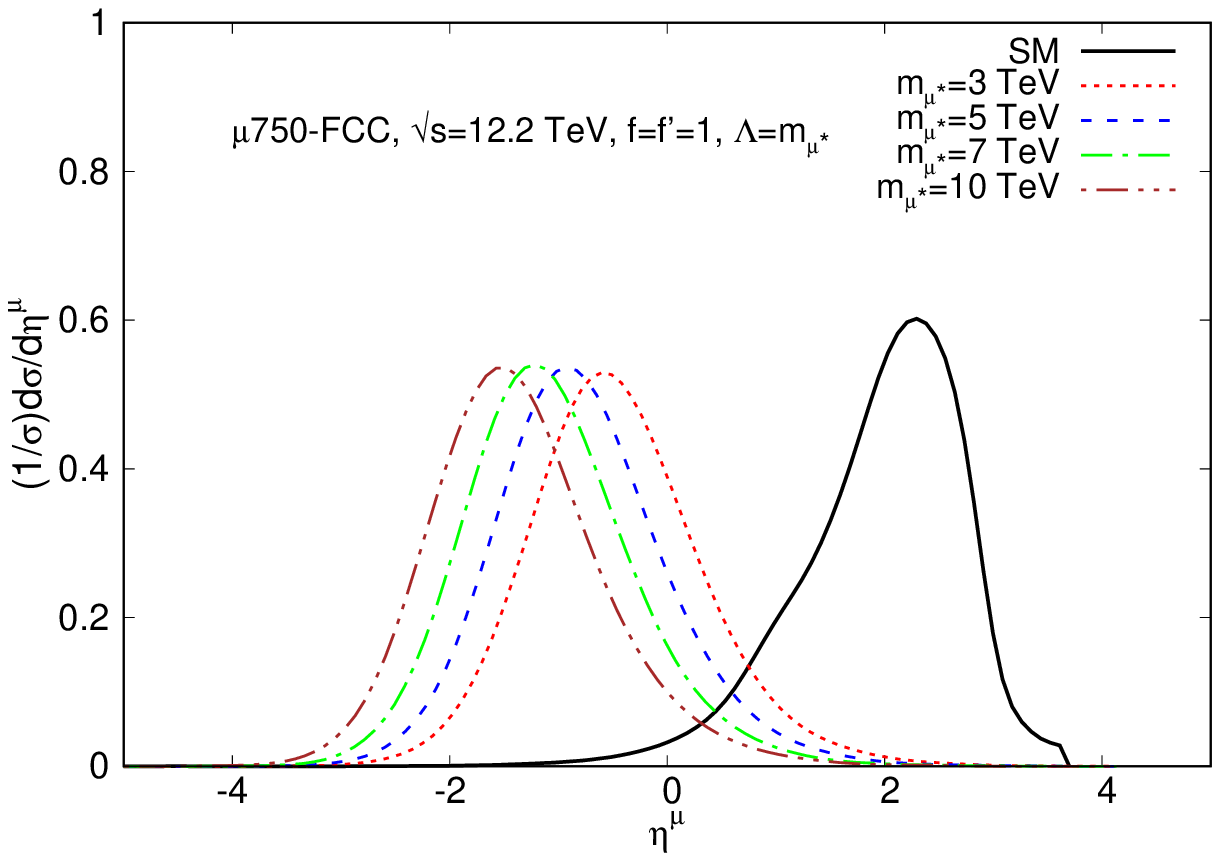}\includegraphics[scale=0.65]{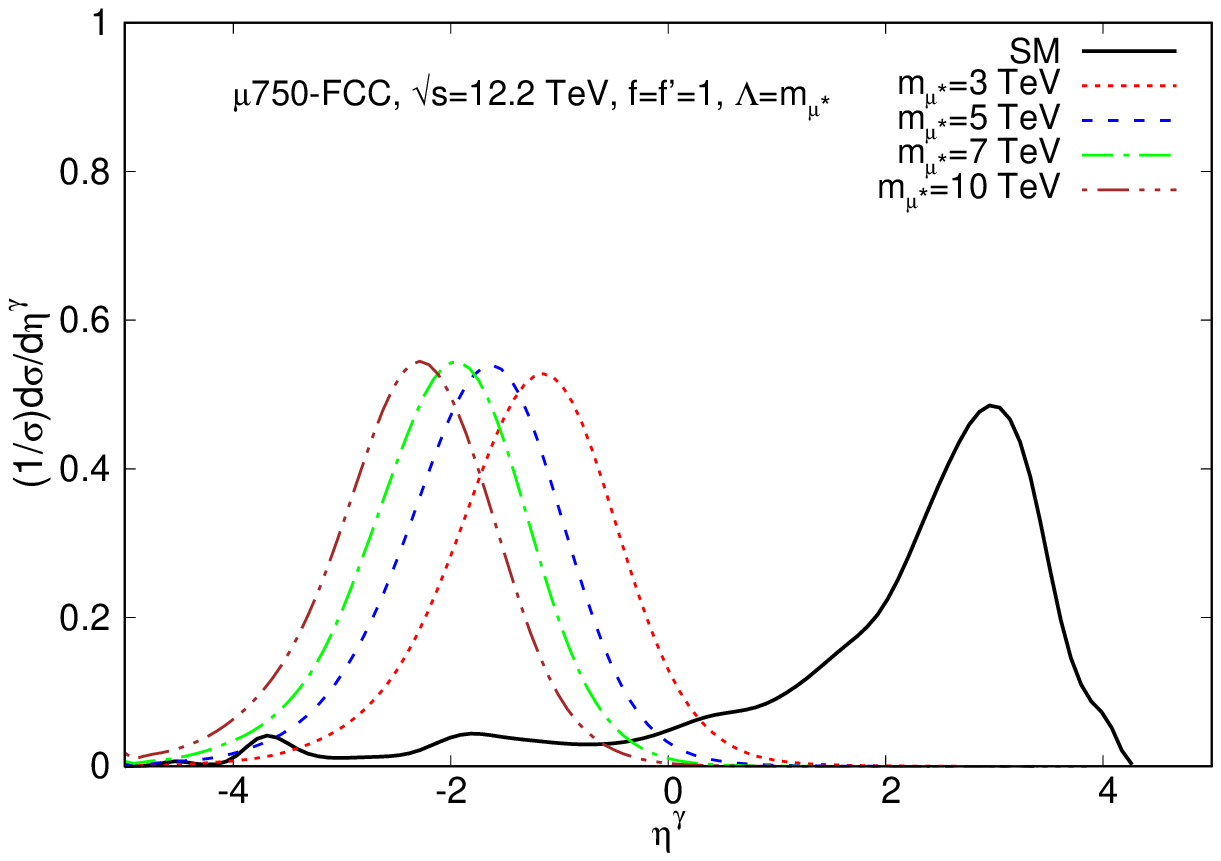}

\caption{Muon (left) and photon (right) normalized $\eta$ distributions for
the $\mu750$-FCC.}
\end{figure}

\begin{figure}[H]
\includegraphics[scale=0.65]{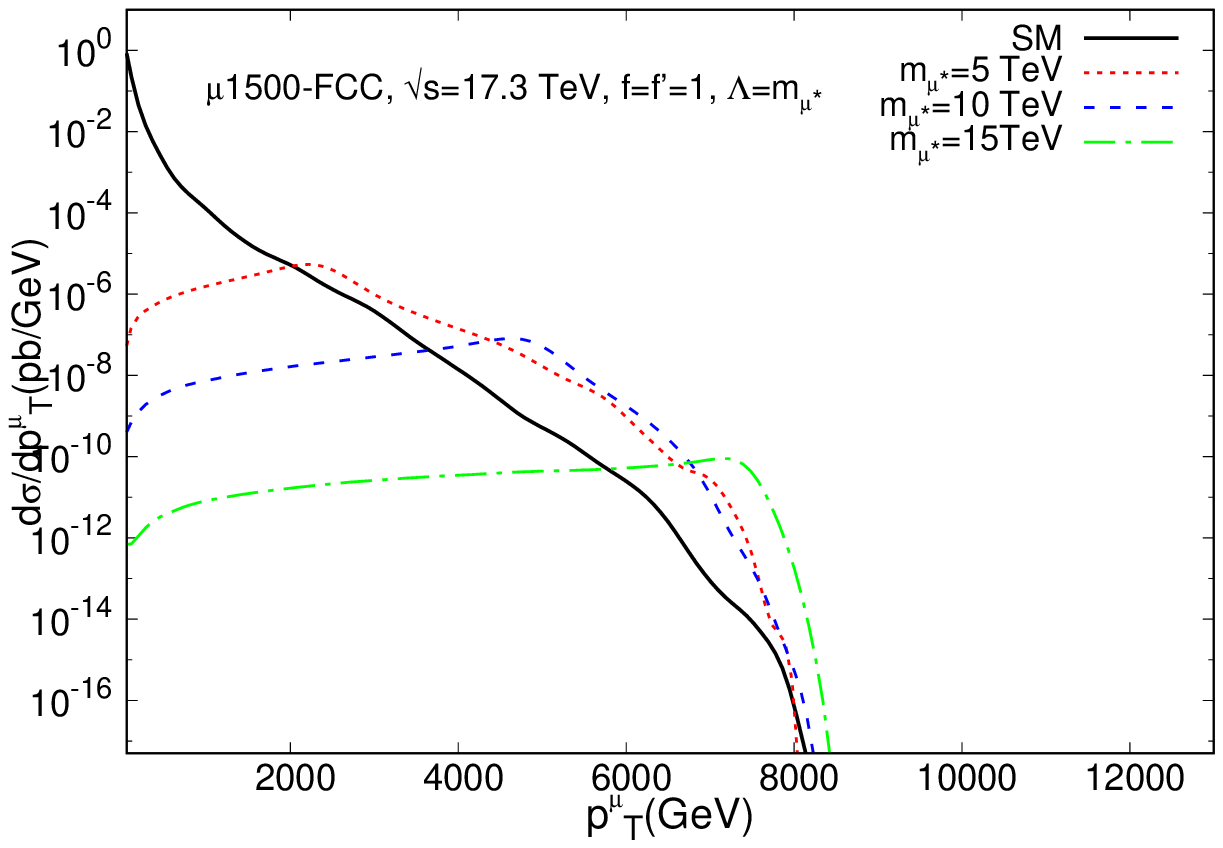}\includegraphics[scale=0.65]{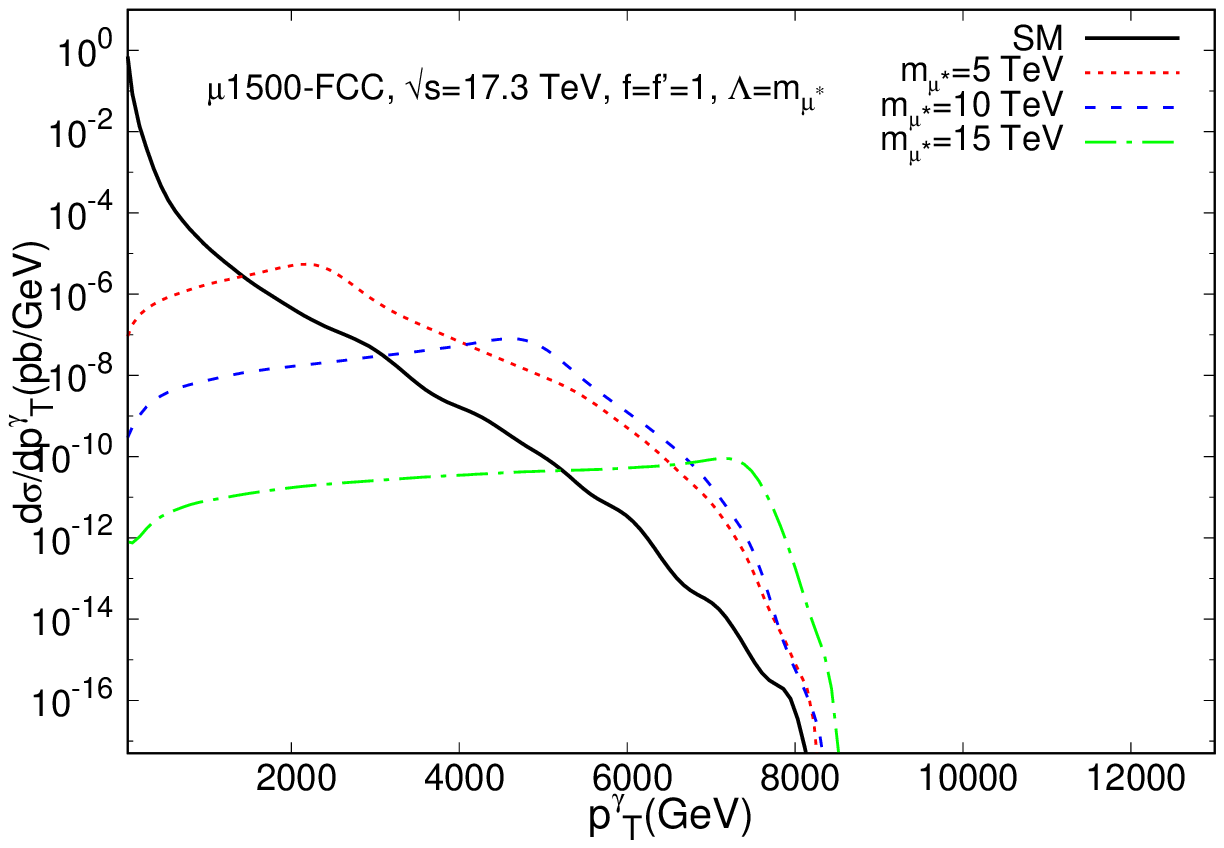}

\caption{Muon (left) and photon (right) $p_{T}$ distributions for the $\mu1500$-FCC.}
\end{figure}

\begin{figure}[H]
\includegraphics[scale=0.65]{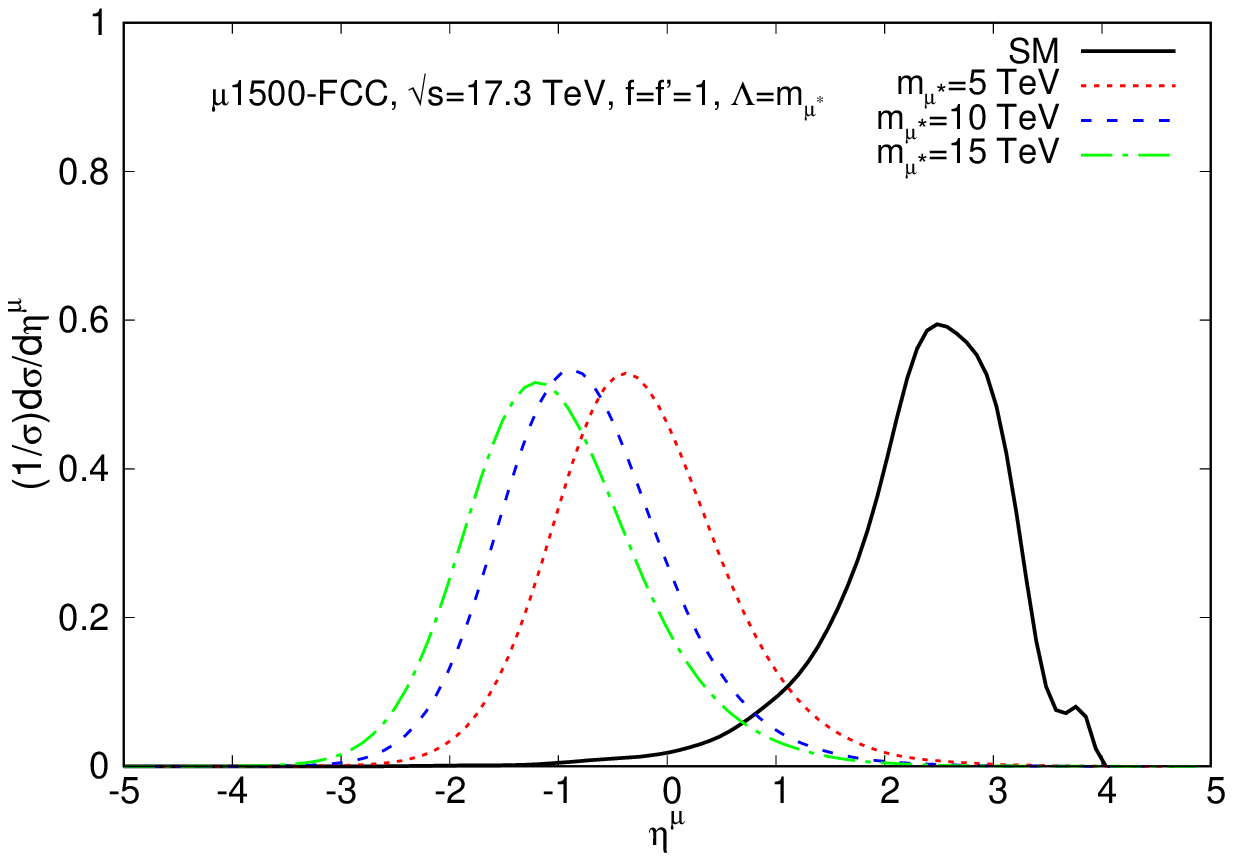}\includegraphics[scale=0.65]{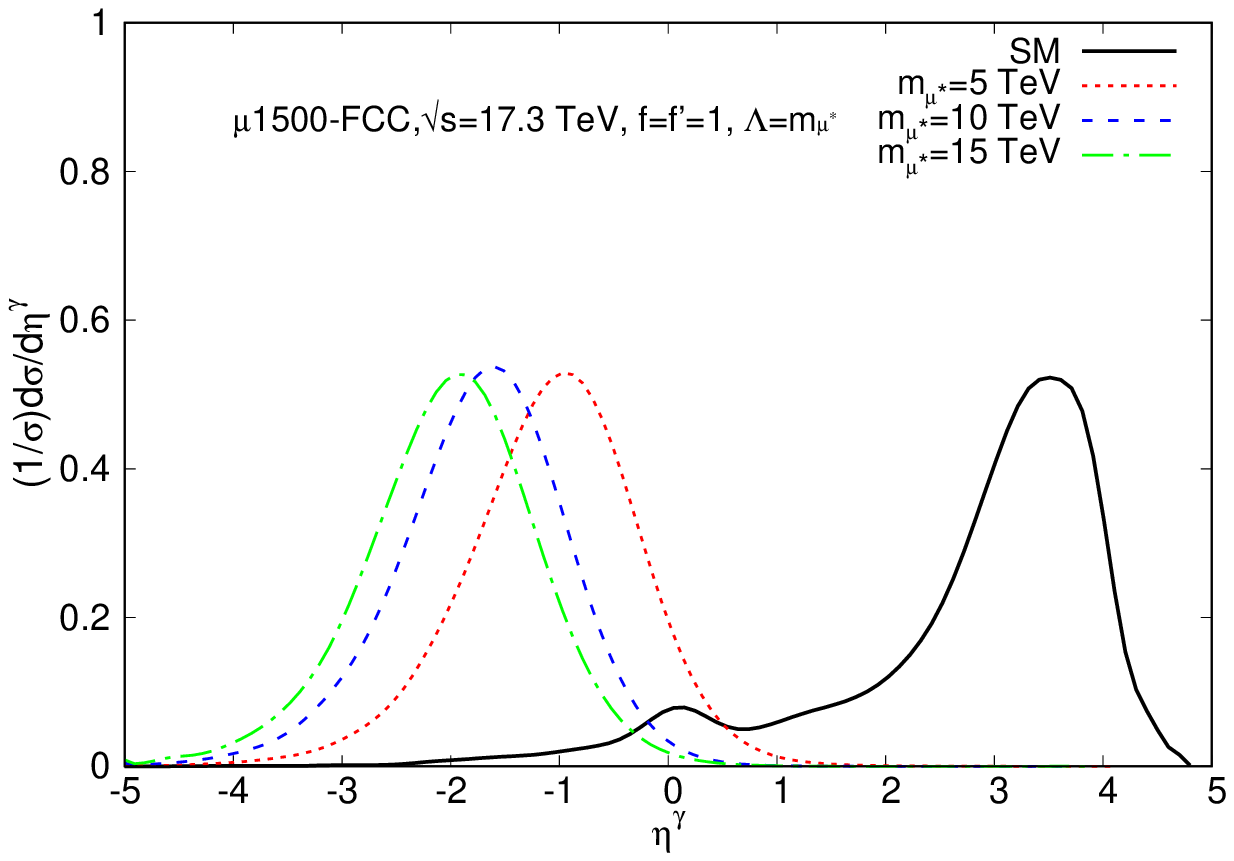}

\caption{Muon (left) and photon (right) normalized $\eta$ distributions for
the $\mu1500$-FCC.}
\end{figure}

\begin{figure}[H]
\begin{centering}
\includegraphics[scale=0.7]{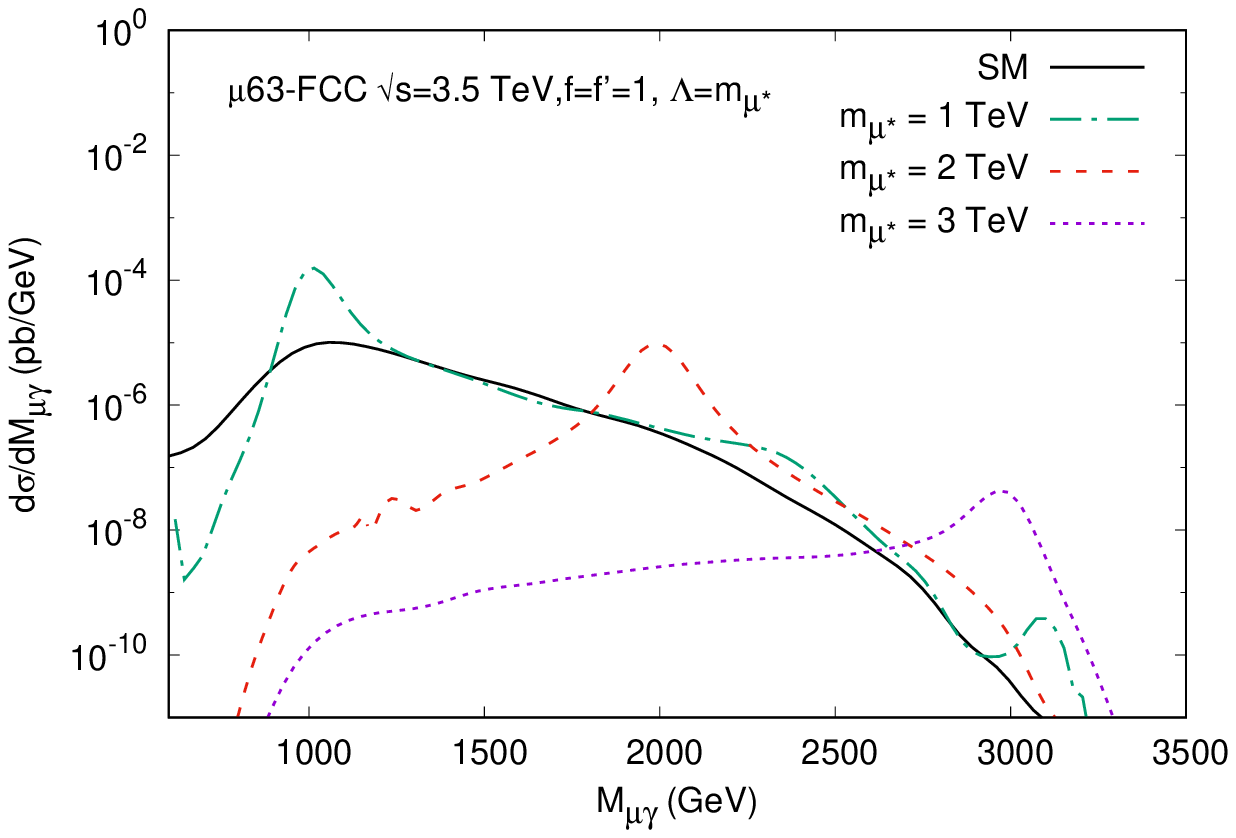}\includegraphics[scale=0.7]{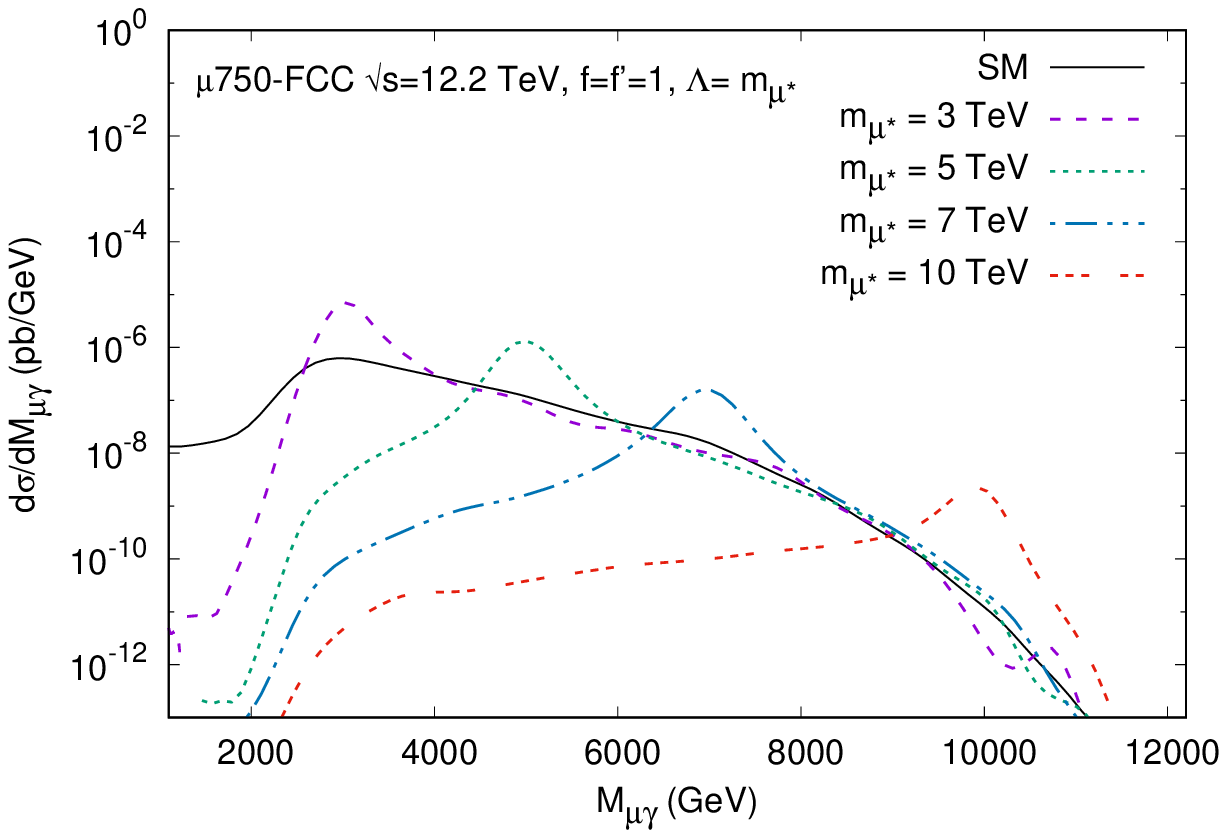}
\par\end{centering}
\begin{centering}
\includegraphics[scale=0.7]{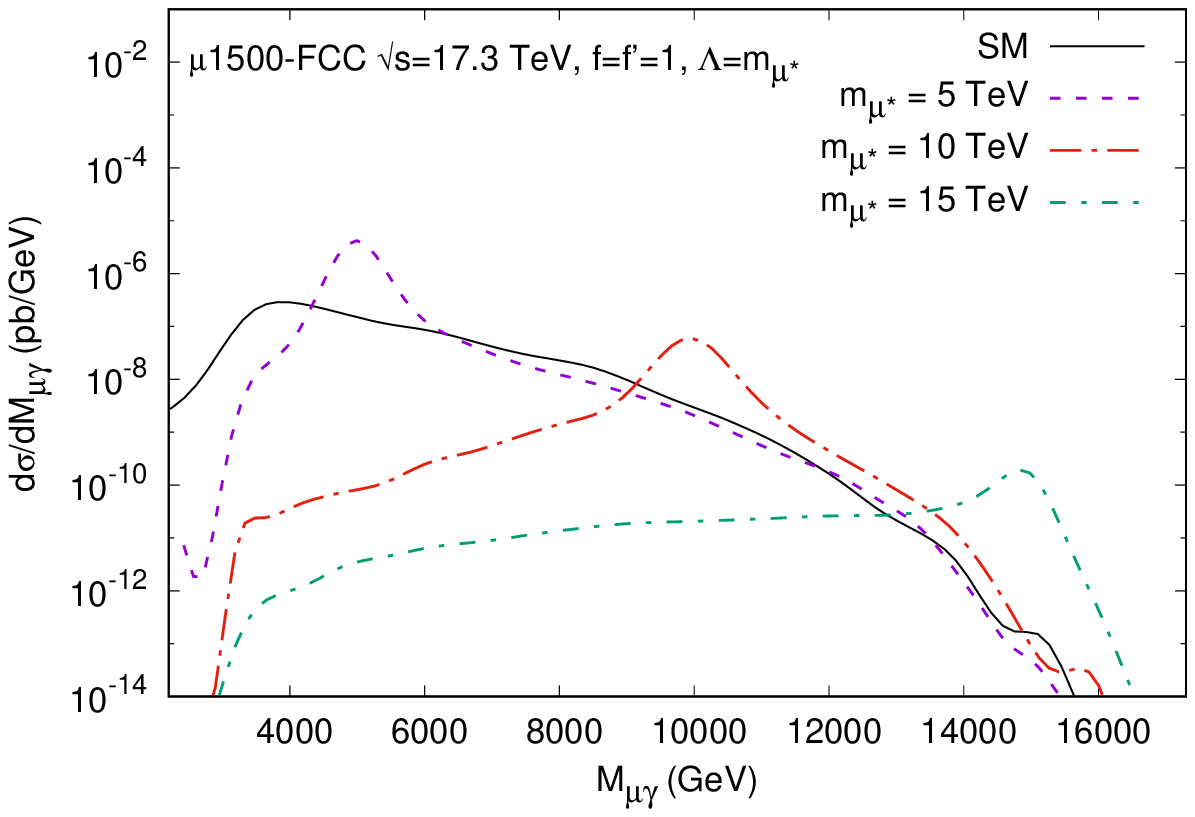}
\par\end{centering}
\caption{Invariant mass distributions of the $\mu\gamma$ system after the
discovery cuts for $\mu63$-FCC, $\mu750$-FCC, and $\mu1500$-FCC,
respectively. }
\end{figure}

By examining these distributions we determine the discovery cuts presented
in Table 2. To determine these discovery cuts we specify the optimal
regions where we cut off the most of the background but at the same
time do not affect the signal so much. Since we choose the $\mu^{\star}\rightarrow\mu\gamma$
decay mode of the excited muon (try to identify the excited muons
through its decay products), no further cut is made on jets. 

\begin{table}[H]
\caption{Discovery cuts.}
\centering{}%
\begin{tabular}{|c|c|c|c|c|}
\hline 
Collider & $p_{T}^{\mu}$ cut & $p_{T}^{\gamma}$ cut & $\eta^{\mu}$ cut & $\eta^{\gamma}$ cut\tabularnewline
\hline 
\hline 
$\mu63$-FCC & $p_{T}^{\mu}>450$ GeV  & $p_{T}^{\gamma}>300$ GeV & $-4.5<\eta^{\mu}<-0.8$ & $-4.8<\eta^{\gamma}<-1.2$\tabularnewline
\hline 
$\mu750$-FCC & $p_{T}^{\mu}>1200$ GeV  & $p_{T}^{\gamma}>900$ GeV & $-3.5<\eta^{\mu}<0.5$ & $-4<\eta^{\gamma}<0.3$\tabularnewline
\hline 
$\mu1500$-FCC & $p_{T}^{\mu}>1500$ GeV  & $p_{T}^{\gamma}>1500$ GeV & $-3<\eta^{\mu}<1$ & $-4<\eta^{\gamma}<0.5$\tabularnewline
\hline 
\end{tabular}
\end{table}

The invariant mass distributions following these cuts are shown in
Figure 10. We define the statistical significance of the expected
signal yield as

\begin{equation}
SS=\frac{\sigma_{S}}{\sqrt{\sigma_{B}}}\sqrt{\epsilon.L_{int}},
\end{equation}
where $\sigma_{S}$ denotes cross-section due to the excited muon
production and $\sigma_{B}$ denotes the SM cross-section, $L_{int}$
is the integrated luminosity of the collider, and $\epsilon$ is the
selection efficiency to detect the signal in the chosen channel ($\epsilon$
is assumed to be the same both on signal and on background). Taking
into account the criteria $SS>3$ (95\% CL) and $SS>5$ (99\% CL),
we derive the mass limits for excited muons. Our results are summarized
in Table 3. 

\begin{table}[H]

\caption{Mass limits for $\mu^{\star}$ at FCC-based $\mu p$ colliders. }
\centering{}%
\begin{tabular}{|c|c|c|c|c|}
\hline 
\multirow{2}{*}{Collider} & \multirow{2}{*}{$L_{\mu p}\,(cm^{-2}s^{-1})$} & \multirow{2}{*}{$\Lambda$} & \multicolumn{2}{c|}{$m_{\mu^{\star}}$(GeV)}\tabularnewline
\cline{4-5} 
 &  &  & $3\sigma$ & $5\sigma$\tabularnewline
\hline 
\multirow{2}{*}{$\mu63$-FCC} & \multirow{2}{*}{$0.2\times10^{31}$} & $m_{\mu^{\star}}$ & 2330 & 2250\tabularnewline
\cline{3-5} 
 &  & $100$ TeV & 2300 & 2180\tabularnewline
\hline 
\multirow{2}{*}{$\mu750$-FCC} & \multirow{2}{*}{$50\times10^{31}$} & $m_{\mu^{\star}}$ & 6500 & 5950\tabularnewline
\cline{3-5} 
 &  & $100$ TeV & 6000 & 5830\tabularnewline
\hline 
\multirow{2}{*}{$\mu1500$-FCC} & \multirow{2}{*}{$50\times10^{31}$} & $m_{\mu^{\star}}$ & 8050 & 7540\tabularnewline
\cline{3-5} 
 &  & $100$ TeV & 7930 & 7480\tabularnewline
\hline 
\end{tabular}
\end{table}

\section{CONCLUSION}

It is shown that the FCC-based muon-proton colliders have a significant
potential in excited muon investigations. We have studied the excited
muon production and decay in various FCC-based $\mu p$ collider options
with muon energies of $63$, $750$, and $1500$ GeV. Our analysis
shows that taking into account the $SS>5$ criteria, for $\varLambda=m^{*}$,
excited muon mass limits are $2250$ GeV, $5950$ GeV, and $7540$
GeV, for $\sqrt{s}=3.5$, $12.2$, and $17.3$ TeV, respectively.
Also, for the same criteria, for $\varLambda=100$ TeV, excited muon
mass limits are 2180, 5830, and 7480 GeV for $\sqrt{s}=3.5$, $12.2$,
and $17.3$ TeV, respectively.
\begin{acknowledgments}
A. Caliskan and S. O. Kara's work is supported by the Scientific and
Technological Research Council of Turkey (TUBITAK) under the Grant
no. 114F337.
\end{acknowledgments}

\end{document}